\documentclass[final,twocolumn]{elsarticle}

\usepackage{lineno,hyperref}
\usepackage{braket}
\modulolinenumbers[5]

\journal{Journal of \LaTeX\ Templates}

\begin{document}

\begin{frontmatter}

\title{Time-Dependent Generator Coordinate Method \\ 
for Many-Particle Tunneling}

\author[1]{N. Hasegawa}
\author[2]{K. Hagino}
\author[1,3]{Y. Tanimura}
\address[1]{Department of physics, Tohoku University, Sendai 980-8578, Japan}
\address[2]{Department of physics, Kyoto University, Kyoto 606-8502, Japan}
\address[3]{Graduate Program on Physics for the Universe, Tohoku University, Sendai 980-8578, Japan}

\begin{abstract}
It has been known that the time-dependent Hartree-Fock (TDHF) method, or the 
time-dependent density functional theory (TDDFT), fails to describe many-body 
quantum tunneling. 
We overcome this problem by 
superposing a few 
time-dependent Slater determinants 
with the time-dependent 
generator coordinate method (TDGCM). 
We apply this method to scattering 
of two $\alpha$ particles in one dimension, and demonstrate that 
the TDGCM method yields a finite tunneling probability even at energies 
below the Coulomb barrier, at which the 
tunneling probability is exactly zero 
in the TDHF. This is the first case in which a many-particle 
tunneling is simulated with a microscopic real-time approach. 
\end{abstract}


\end{frontmatter}


\section{Introduction}
One of the primary goals of nuclear reaction theory is to 
develop a microscopic 
framework for nuclear reactions, starting from the nucleonic
degrees of freedom. Such a framework will describe several 
complicated reaction processes in a unified way, not specialized 
in a certain reaction system. 
Ideally, such a framework will not contain any empirical parameter 
for reactions, once static nuclear properties or the 
nucleon-nucleon interaction are well investigated. 
This feature will be particularly important in applying the framework to 
unknown regions, in which experimental studies are difficult, e.g., 
reactions of neutron-rich nuclei and those for superheavy nuclei. 

The time-dependent Hartree-Fock (TDHF) 
method is one of the promising microscopic 
frameworks for nuclear reactions, in which a many-body wave function 
is approximated by a single Slater determinant 
\cite{RS80,BKN76,Negele82,Simenel12,Nakatsukasa16}. 
The TDHF has been successful in describing average behaviors 
of nuclear reactions, such as the energy-angle correlation 
in heavy-ion deep inelastic collisions \cite{Dhar81}. 
In recent years, many TDHF calculations have been 
successfully carried out with current powerful 
computers, in order to discuss the main process of reaction for a given 
initial condition 
\cite{Sky3D,Umar15,Golabek09,Sekizawa13,Sekizawa16,Washiyama08,Scamps13,Bulgac16,patterson2019photoabsorption}. 

However, it has been well known that the TDHF has a serious drawback, that 
is, it considerably underestimates quantum fluctuations 
and fails to 
describe minor processes, such as quantum many-body tunneling at energies below 
the potential barrier \cite{Negele82,Simenel12,Nakatsukasa16}. 
Notice that the TDHF can be formulated using the 
stationary phase approximation to the path integral representation of a 
many-body dynamics \cite{Negele82}, and in this sense 
the collective path in the TDHF is classical. A tunneling probability 
in the TDHF therefore changes abruptly from 0 to 1 at a 
certain threshold energy. For this reason, the TDHF method fails to describe 
fusion reactions at energies below the barrier \cite{Simenel12} 
and fission dynamics before a saddle point \cite{Tanimura17}. 

Several schemes have been considered so far in order to go beyond the 
TDHF approximation 
\cite{Wong78,WC85,GT90,ayik2008stochastic,LA14,colonna1994unstable,reinhard1983time,Reinhard92,Lacombe16,Grange81,Reinhardt84,Balian84}.
Nevertheless none of them has been 
applied to the problem of quantum tunneling. 
For instance, in the stochastic mean-field theory (SMF) 
\cite{Tanimura17,ayik2008stochastic,LA14}, 
quantum fluctuations are simulated as an ensemble of TDHF trajectories, 
but it is still difficult to describe 
many-body tunneling with this framework.

Since a TDHF trajectory shows a classical behavior, 
to describe quantum tunneling based on the TDHF and its extension 
has a common feature to a problem of how to simulate quantum tunneling 
using classical trajectories. This problem has been actually discussed also in 
the field of quantum chemistry, in which 
the entangled trajectory
molecular dynamics (ETMD) has been developed 
\cite{DM01,wang2009quantum,wang12,xu2016quantum}. 
In this method, the Winger function of a one particle wave function is 
represented as an ensemble of classical test particles. 
Those classical test particles are treated in a collective manner, rather than 
independently. By taking into account such entanglement of test particles, 
that is, the influence of other test 
particles during the time evolution of a test particle, it has been 
demonstrated that the ETMD can simulate well 
quantum tunneling phenomena, despite that each test particle follows 
the classical equation of motion 
\cite{DM01,wang2009quantum,wang12,xu2016quantum}. 

The success of ETMD would imply that quantum tunneling may be simulated 
by taking into account an entanglement of many TDHF trajectories. 
This would be equivalent to construct a many-body wave function as 
a superposition of many Slater determinants. This is nothing but the 
time-dependent generator coordinate method (TDGCM) 
\cite{reinhard1983time,Orestes07}. 
The aim of this paper is to 
investigate whether the TDGCM method can 
describe in principle quantum tunneling phenomena. 
For such a proof of principle study, it would be ideal to choose 
a system which is as simple as possible. To this end,  
we consider a collision of two $^4$He particles in the one dimensional space. 
Based on the idea of ETMD, 
we superpose a few Slater determinants, which have different 
initial relative distances 
and relative momenta between the two $^4$He particles. 
During the time evolution, the trajectory of each Slater determinant 
is affected by the presence of other 
Slater determinants, and they are thus entangled as in the entangled 
classical trajectories in the ETMD method. 
We shall investigate in this paper 
whether such entanglement of Slater determinants 
improves the failure of the TDHF method. 

\section{Time-dependent Generator Coordinate Method}

In the TDGCM, one assumes that a many-body wave function is given as 
a superposition of many Slater determinants, 
\begin{equation}
 \Psi(t)=\sum_a f_a(t)\Phi_a(t), 
\label{GCMwf}
\end{equation}
where $f_a$ is a weight function and 
$\Phi_a$ is a Slater determinant with single-particle wave functions 
$\{\phi_{ai}\}$. 
The index $a$ distinguishes each Slater determinant to one another, and 
is referred to as a generator coordinate. In nuclear structure calculations, 
multipole moments are often taken as a generator coordinate, for which the 
Slater determinant $\Phi_a$ is obtained with the constrained 
Hartree-Fock method \cite{RS80}. 
In the calculation shown below, we take the initial 
values of the relative distance and the relative momentum between two 
$\alpha$ particles as the generator coordinates. 

The time evolution of the weight functions $f_a(t)$ and the Slater 
determinants $\Phi_a(t)$ can be determined by the 
time-dependent variational principle, 
\begin{equation}
 \delta\int dt \frac{\braket{\Psi|i\hbar\frac{\partial}{\partial t}-H|\Psi}}{\braket{\Psi|\Psi}}=0,
\label{154810_12Nov19}
\end{equation}
where $H$ is the total Hamiltonian of the system. 
In Refs. \cite{reinhard1983time,Orestes07,Regnier19}, the time evolution of 
the Slater determinants is replaced by TDHF trajectories, while the 
time evolution of the weight function is solved according to the variational 
principle. However, in an application to quantum tunneling, it would be 
essential to take into account a deviation from TDHF 
in the evolution of the Slater determinants. 
We therefore do not use TDHF trajectories, but determine the time-evolution 
of the Slater determinants 
under the influence of the other Slater determinants 
using the variational principle. 

Notice that 
our method is conceptually different from the method developed in 
Refs. 
\cite{zdeb2017fission,regnier2016fission,bernard2011microscopic,goutte2005microscopic,tao2017microscopic}, even though 
the latter has often been called the time-dependent generator 
coordinate method. In Refs. 
\cite{zdeb2017fission,regnier2016fission,bernard2011microscopic,goutte2005microscopic,tao2017microscopic}, the Slater determinants $\Phi_a$ are prefixed 
by nuclear structure calculations, often taken as the ground state for a given 
value of multipole moments, whereas the time evolution of the weight functions 
is determined by solving the time-dependent Hill-Wheeler equation. Since the 
Slater determinants are static in these calculations, it may be more 
appropriate to call such method as the time-dependent Hill-Wheeler method, 
rather than the time-dependent generator coordinate method. 
Even though non-adiabatic effects can in principle be taken into account 
in this method \cite{bernard2011microscopic}, the computational cost 
considerably increases when the number of configurations increases. 
In contrast, the non-adiabatic effects are automatically taken into 
account in our method by using time-dependent Slater determinants.  
This would be an important aspect of our method, especially in applications 
to nuclear reactions of heavy systems, in which excitations of colliding 
nuclei during reaction play an important role \cite{HT12}. 

\section{Application to $\alpha$+$\alpha$ scattering in one dimension}

Let us now apply the TDGCM to the $^4$He+$^4$He scattering in one dimension. 
In order to simplify the calculation, we parameterize the single-particle 
wave functions with a Gaussian function with a fixed width, as in the 
anti-symmetrized molecular dynamics (AMD) 
\cite{En'yo12,ono1992antisymmetrized}. 
That is, we take the single-particle wave function $\phi_{ai}$ as, 
\begin{equation}
\phi_{ai}(x,t) = e^{-\nu\left(x-\frac{Z_{ai}(t)}{\sqrt{\nu}}\right)^2}\chi_\sigma\chi_\tau, 
\end{equation}
where $\chi_\sigma$ and $\chi_\tau$ are the spin and the isospin wave 
functions, respectively.  
The Gaussian center, $Z_{ai}$, is a complex 
quantity, whose real and imaginary parts are related to the mean position 
and the mean momentum of the Gaussian wave function, respectively.

In this paper, we fix the spin and isospin wave functions
and assume that the Gaussian center is identical 
for all the four nucleons in each $^4$He particle. 
That is, we take 
$Z_{p\uparrow}=Z_{p\downarrow}=Z_{n\uparrow}=Z_{n\downarrow}$ in each $^4$He particle. 
Moreover, we assume that there is no transfer of nucleons during a reaction, 
and the two $^4$He move symmetrically with respect to $x=0$. That is, 
when one $^4$He particle has a Gaussian center of $Z(t)$, the other 
$^4$He particle has a Gaussian center of $-Z(t)$. 
With these approximations, there is only one single variational parameter, 
$Z_a(t)$, in each Slater determinant, $\Phi_a(t)$. 

At $t=0$, we construct 
each Slater determinant, $\Phi_a(t)$, by putting 
the two $^4$He particles at $x=x_{a0}$ and $x=-x_{a0}$, with 
the momentum of $p=-p_{a0}$ and $p=p_{a0}$, respectively. 
This is equivalent to take 
\begin{equation}
Z_a(t=0)=x_{a0}\sqrt{\nu} -i \frac{p_{a0}}{2\sqrt{\nu}\hbar}, 
\end{equation}
for a given Gaussian width, $\nu$. 

We employ the same one dimensional Hamiltonian as 
the one in Ref. \cite{reinhard1983time}, to which we also add 
a soft-core Coulomb interaction between two protons in a 
form of \cite{Pindzola91,Grobe92,Grobe93,Lein00,Maruyama12}
\begin{equation}
v_C(x,x')=\frac{e^2}{\sqrt{\alpha^2+(x-x')^2}}.
\end{equation}
With this Hamiltonian, one obtains \cite{Flocard76}, 
\begin{eqnarray}
&&\braket{\Phi_a|H|\Phi_{a'}} \nonumber \\
&&=
\braket{\Phi_a|\Phi_{a'}}
\int dx\left\{\frac{\hbar}{2m}\tilde{\tau}(x)
+\frac{t_3}{3}\tilde{\rho}^3(x)\right.
\nonumber \\
&&+\frac{t_0}{2}\tilde{\rho}(x)\int dx'\tilde{\rho}(x')
\frac{b}{\sqrt{\pi}}\,e^{-\frac{(x-x')^2}{b^2}} \nonumber \\
&&\left.+\frac{e^2}{2}\tilde{\rho_p}(x)\int dx'\tilde{\rho_p}(x')
\frac{1}{\sqrt{\alpha^2+(x-x')^2}}\right\}, 
\label{165719_3Dec19}
\end{eqnarray}
with
\begin{eqnarray}
\tilde{\tau}(x)&=&\sum_{i,j}\left(\frac{\partial}{\partial x}\phi_{ai}^*(x)\right)
\left(\frac{\partial}{\partial x}\phi_{a'j}(x)\right)
\left(B^{-1}\right)_{ji}\nonumber \\
\label{165752_3Dec19}\\
 \tilde{\rho}(x)&=&\sum_{i,j}\phi_{ai}^*(x)\phi_{a'j}(x)\left(B^{-1}\right)_{ji},
\label{165756_3Dec19}
\end{eqnarray}
where the matrix $B$ is defined as 
$B_{ij}=\braket{\phi_{ai}|\phi_{a'j}}$. 
The overlap kernel $\braket{\Phi_a|\Phi_{a'}}$ in Eq. (\ref{165719_3Dec19}) 
is given as $\braket{\Phi_a|\Phi_{a'}}={\rm det}(B)$. 
We follow Refs. \cite{reinhard1983time,Richert79} and use 
the parameters of 
$t_0=-12.5$ MeV fm$^{-1}$, $t_3=8.8$ MeV fm$^{-2}$, and $b=2.0$ fm.
For the Coulomb interaction, we arbitrarily set $\alpha=1.0$ fm. 

In order to derive the time-dependent equation for $Z_a(t)$ according 
to the variational principle, Eq. (\ref{154810_12Nov19}), 
one needs to evaluate the derivative of 
$\langle \Psi|H|\Psi\rangle/\langle \Psi|\Psi\rangle$ with respect to 
$Z^*_a(t)$. We carry it out numerically, that is, we calculate a change 
of this quantity when $Z^*_a(t)$ is shifted to $Z^*_a(t)+\delta Z^*_a(t)$ 
and divide it by $\delta Z^*_a(t)$. 
Notice that
the derivative 
does not depend on the direction of $\delta Z^*_a(t)$ in the complex plane, 
as long as the absolute value of $\delta Z^*_a(t)$ is small. 

In the calculations shown below, we use the Gaussian width 
of $\nu=0.5$ fm$^{-2}$. This value is determined so that 
the internucleus potential at $x=0$ evaluated in the frozen density 
approximation with a single Slater determinant \cite{Washiyama08} 
becomes higher than the 
height of the Coulomb barrier. With the parameter set employed, 
the barrier in the frozen density approximation 
is located at 3.84 fm with the height of 0.13 MeV, while 
the potential at $x=0$ is 7.1 MeV. 


\begin{figure}
\includegraphics[width=8cm]{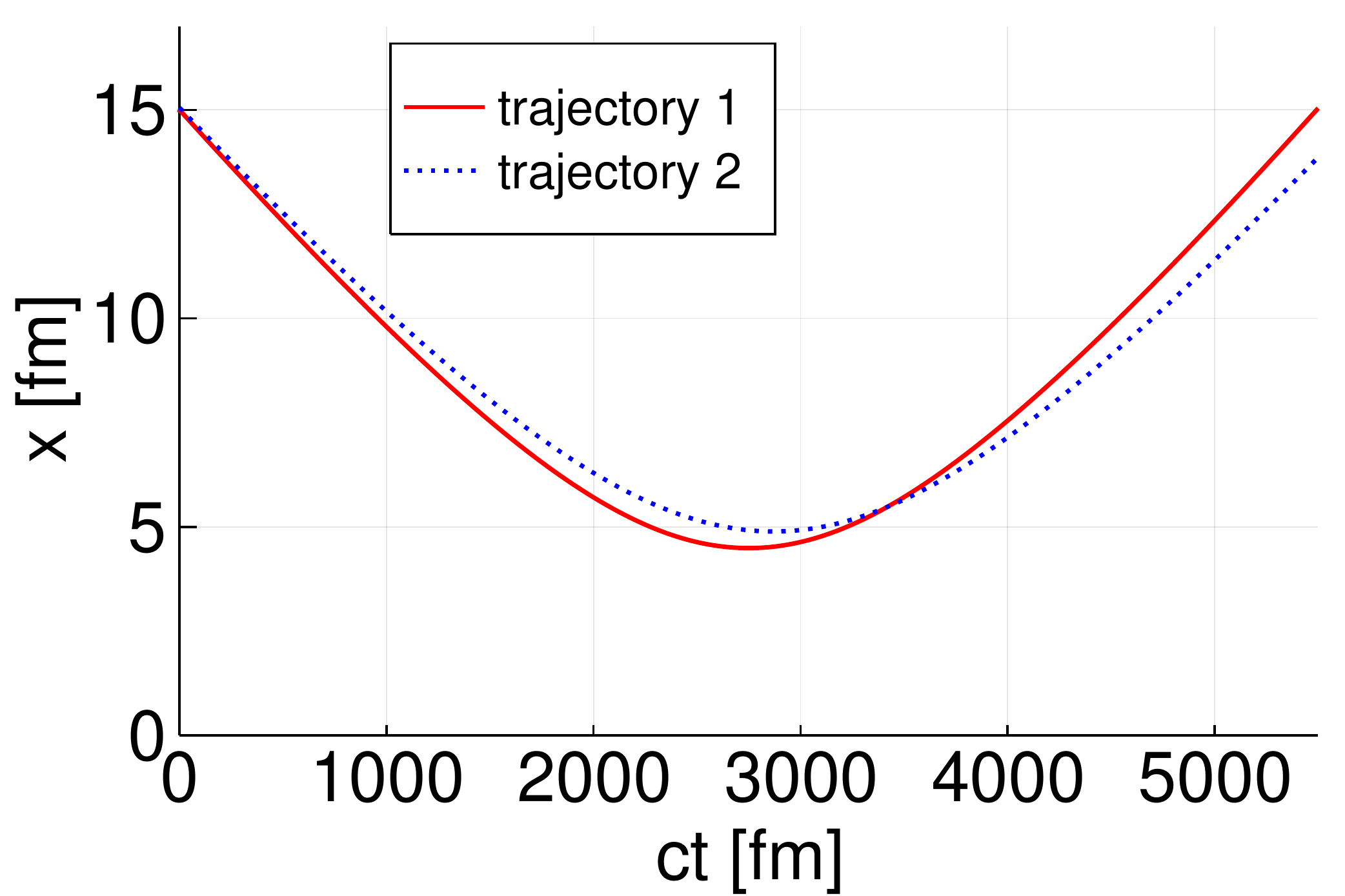}
\caption{
The trajectory of the right $\alpha$ particle in $\alpha$+$\alpha$ 
scattering in one dimension 
obtained with the time-dependent Hartree-Fock with 
two different initial conditions. The two $\alpha$ particles are 
assumed to move symmetrically with respect to $x=0$. 
The solid line is obtained with the initial condition of 
$x_0$=30.0/2 fm and $E=$0.113 MeV, while the dashed line is obtained with 
$x_0$=30.1/2 fm and $E=$0.100 MeV. The interaction is based on a 
Gaussian + density-dependent zero-range
nuclear interaction together with a soft-core Coulomb interaction. 
The Coulomb barrier between the two $\alpha$ particles is located at 
$x=3.84$ fm with the height of 0.13 MeV. }
 \label{123437_28Nov19}
\end{figure}

\begin{figure}
 \includegraphics[width=8cm]{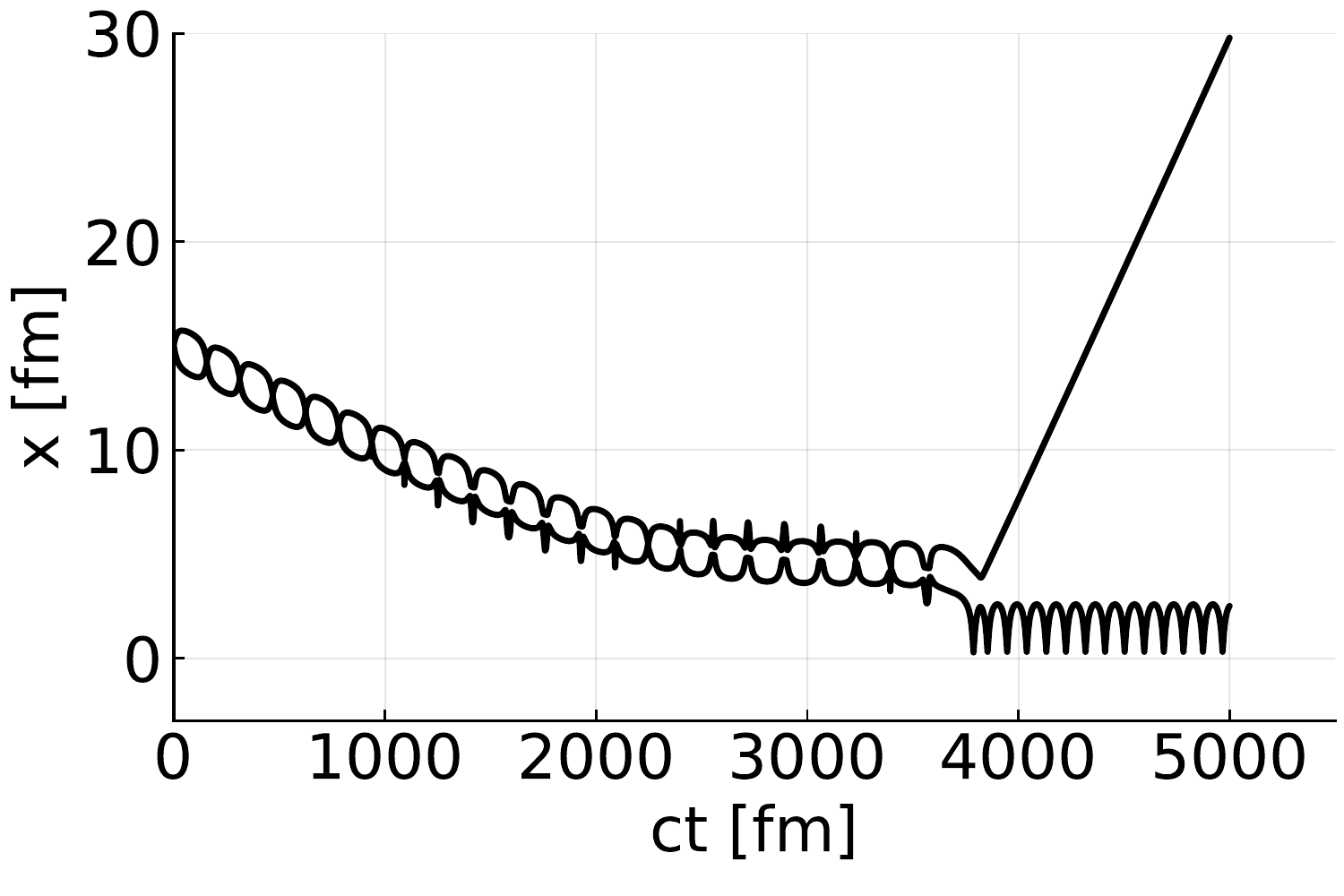}
 \caption{
Same as Fig. \ref{123437_28Nov19}, but obtained with the 
time-dependent generator coordinate method with the same initial 
conditions. }
 \label{205817_28Nov19}
\end{figure}

We first discuss the result of a superposition of two Slater determinants. 
For the first Slater determinant, the initial positions of the $\alpha$ 
particles are set to be $x_0=\pm 30.0/2$ fm with the initial relative momentum 
of $p_0=\mp \sqrt{2\mu E}$ with $E=0.113$ MeV, where $\mu$ is the reduced mass 
evaluated with the static calculation for a single $\alpha$ particle, 
whereas for the second Slater determinant we take $x_0=\pm 30.1/2$ fm and 
$p_0=\mp \sqrt{2\mu E}$ with $E=0.100$ MeV. 
Fig. \ref{123437_28Nov19} shows the position of each 
Gaussian wave packet, $x(t)={\rm Re}[Z(t)]/\sqrt{\nu}$, where ${\rm Re}[Z]$ 
denotes the real part of $Z$, for the case 
where the Slater determinants are evolved 
independently to each other with the TDHF method. 
Since the initial energy is below the barrier for both the 
Slater determinants, 
they are both reflected at the barrier located at 3.84 fm. 

Figure \ref{205817_28Nov19} shows the result of the TDGCM with two Slater 
determinants, for which the initial conditions are taken to be the same as 
those in the independent TDHF calculations in Fig. \ref{123437_28Nov19}. 
In this calculation, we initially take equal weights for the two Slater determinants, that is, 
$f_1=f_2$.

The average initial relative energy,  
$E_{\rm rel}\equiv\langle \Psi|H|\Psi\rangle/\langle \Psi|\Psi\rangle-2E_{\rm g.s.}$, 
where $E_{\rm g.s.}$ is the ground state energy of the $\alpha$ particle, 
is calculated to be 
0.099 MeV, which is still below the Coulomb barrier. 
Yet, one can clearly see that 
one of the trajectories overcomes the barrier and undergoes the fusion 
process, while the other trajectory is reflected at the barrier. 
This is in marked contrast to the TDHF case shown in 
Fig. \ref{123437_28Nov19}, in which both of the trajectories 
are reflected at the barrier. 
That is, the tunneling probability of the Coulomb barrier is zero in the TDHF, 
but it is finite in the TDGCM, even though the exact value of the tunneling 
probability is difficult to evaluate only with two Slater determinants.  
\begin{figure}
 \includegraphics[width=8cm]{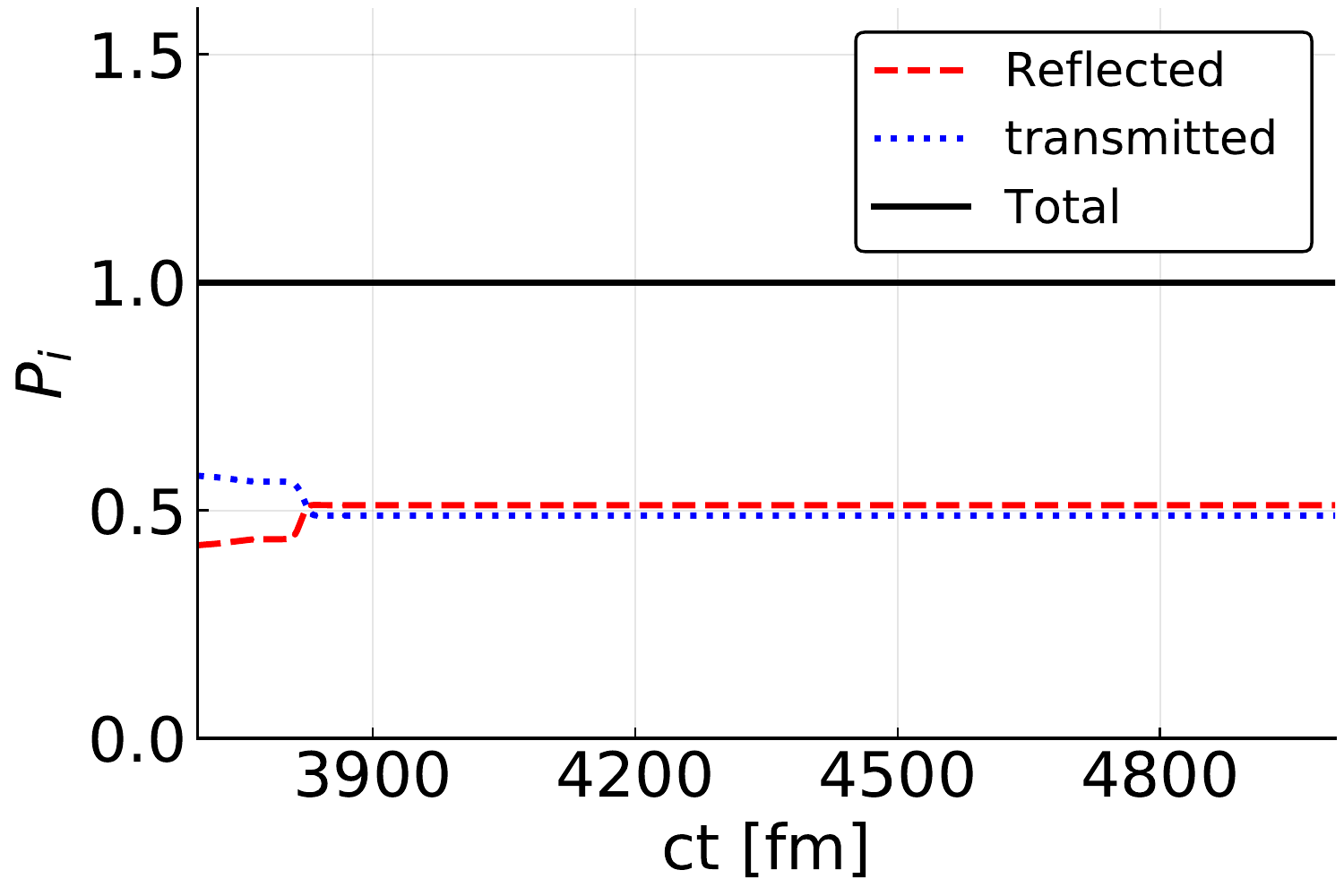}
 \caption{The time-evolution of the fractions of each Slater determinant shown in Fig. 2.
 The dashed and the dotted lines show the reflected and the transmitted Slater determinant respectively.}
 \label{180905_19May20}
\end{figure}
Figure \ref{180905_19May20} shows the time-evolution of
the probabilities of each trajectory in the wave function.
Here, the probabilities 
are defined as (see Eq. (\ref{GCMwf})), 
\begin{equation}
 P_i(t)=\frac{|f_i(t)|^2\braket{\Phi_i(t)|\Phi_i(t)}}{\braket{\Psi(t)|\Psi(t)}}. 
\end{equation}
Notice that 
we only plot the long-time behavior, for which the overlap between the two Slater determinant is small, 
that is, $\braket{\Phi_1(t)|\Phi_2(t)}\sim 0$. 
The fraction for the transmitted trajectory becomes from 0.5 at $t=0$ to 0.489 at $t=\infty$. 
This value might be identified with the tunneling probability.

In the case of ETMD, 
a finite tunneling probability is explained as that 
the energy of individual test particles is not conserved, and 
some of test particles can ``borrow'' an energy from an ensemble of 
the other test particles in order to overcome the barrier 
\cite{DM01,wang2009quantum,wang12,xu2016quantum}. 
We anticipate that the same argument applies to the TDGCM as well, 
even though it is difficult to visualize the energy of each 
Slater determinant as a function of time due to the non-orthogonality 
of the Slater determinants. 

\begin{figure}
 \includegraphics[width=8cm]{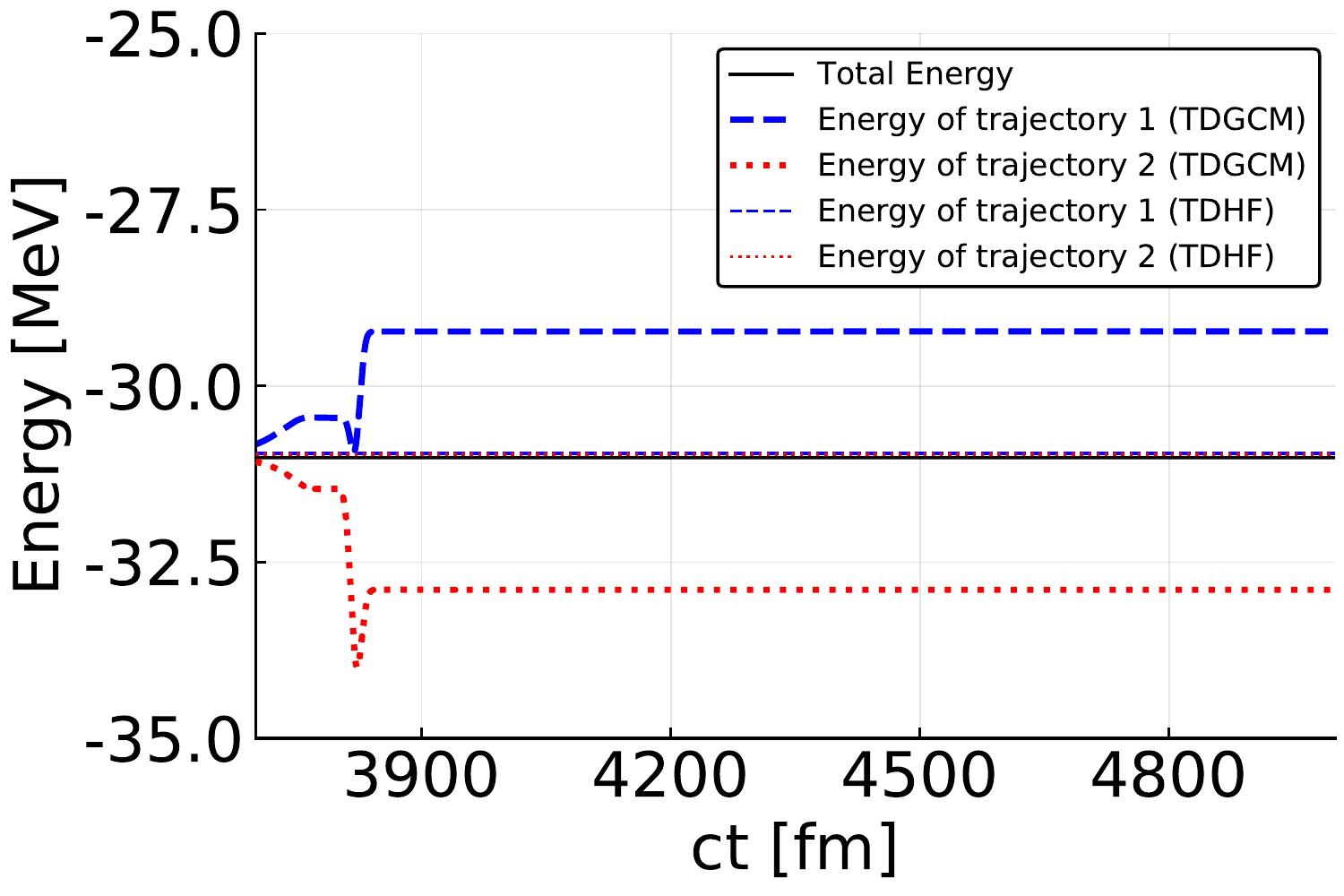}
 \caption{
 The energy of each Slater determinant in the time-dependent generator coordinate method (the thick 
dashed and the dotted lines)
 and that for the TDHF trajectories (the thin dashed line ($-30.95$MeV) and the dotted line ($-30.96$MeV)).
The total energy is also shown by the solid line. While the total energy is conserved, the energy of each trajectory changes as a function of time.
}
 \label{143830_13May20}
\end{figure}

Figure \ref{143830_13May20} shows the behavior of the energy of each trajectory
after the overlap of the Slater determinants becomes appreciably small.
At such $t$, 
we can define the energy of each trajectory as
\begin{equation}
 E_a(t)=\frac{\braket{\Phi_a(t)|H|\Phi_a(t)}}{\braket{\Phi_a(t)|\Phi_a(t)}}\label{205950_10Jun20}
\end{equation}
As compared to the energies for the independent 
Slater determinants (see the thin dashed and dotted lines), 
the energy of one of the trajectories (the trajectory 1; the thick dashed line) increases
while that of the other trajectory (the trajectory 2; the thick dotted line) decreases, 
even though the total energy of the system,  
\begin{equation}
 E_{\rm tot}=\frac{\sum_{ab}f_a(t)^*f_b(t)\braket{\Phi_a(t)|H|\Phi_b(t)}}
{\sum_{ab}f_a(t)^*f_b(t)\braket{\Phi_a(t)|\Phi_b(t)}}, 
\label{210050_10Jun20}
\end{equation}
remains a constant at any time (see the solid line). 
Notice that the trajectory 1 corresponds to the transmitted fraction in Fig. 3, while 
the trajectory 2 corresponds to the reflected fraction in Fig.3. 
This indicates that the trajectory 1 borrows the energy from the trajectory 2 and passes the barrier.

We next discuss the result with 10 Slater determinants. 
To this end, we take the same weight factor $f_a$ for each Slater determinant at $t=0$ 
and randomly generate the initial conditions for $x_{a0}$ and $p_{a0}$ around a central 
value assuming a Gaussian distribution. 
For the initial distances, we take the central value of 30.0/2 fm with 
the Gaussian width of 0.1 fm. On the other hand, for the initial energies, 
we take the central value of 0.1 MeV with the width of 0.02 MeV. 
We use a particular ensemble where all the 10 Slater determinants have an 
energy 
below the barrier so that they are all reflected at the barrier 
when they are evolved individually. 
The average energy of this ensemble is computed to be 0.11 MeV. 
By taking into account the entanglement among the Slater determinants with the 
TDGCM method, we find that 3 Slater determinants undergo fusion, whereas 
the other 7 Slater determinants are reflected at the barrier. 
As in the case of a superposition of 2 Slater determinants, it is 
remarkable that the tunneling phenomenon is simulated using 
``classical'' TDHF trajectories. 
Evidently, the TDGCM provides a promising microscopic framework 
to describe many-particle tunneling phenomena. 

\section{Summary}

In summary, we have applied 
the time-dependent generator coordinate method (TDGCM) 
to the $^4$He+$^4$He scattering in one dimension 
at energies below the Coulomb barrier. 
To this end, we have used the initial relative distance and the initial 
relative momentum between the two $^4$He particles as generator coordinates 
which characterize each Slater determinant. 
We have shown that, by superposing Slater determinants, a many-body 
tunneling can be simulated with this method, whereas the time-dependent 
Hartree-Fock (TDHF) method yields the tunneling probability of either 0 or 1. 
That is, even when individual TDHF trajectories are all reflected at the 
barrier, some of them undergo fusion in the TDGCM. 
It is thus evident that the TDGCM provides a promising means 
to microscopically describe many-body 
tunneling. 

Since the aim of this paper was to carry out a proof-of-principle study 
of the TDGCM framework, for simplicity we 
have parameterized single-particle wave functions in a Gaussian form 
with a fixed width. 
Similar wave functions may be employed in applications of the TDGCM to 
scattering of heavier nuclei in the more realistic three-dimensional space.  
That is, if one assumes that the single-particle wave functions
in the colliding nuclei are given by pre-fixed mean-field potentials, 
those single-particle wave functions are characterized only by the time-dependent centers of the mean-field 
potentials. 
TDGCM calculations can thus be carried out in the same way as what we have 
done in this paper for scattering of two $\alpha$ particle 
in one-dimension. 

Of course, one can improve the performance of the TDGCM 
method by using less restricted single-particle wave functions, even 
though this would be numerically challenging, especially in applications 
to reactions in the three-dimensional space. 
Moreover, we remark that an application of the TDGCM is not restricted 
to nuclear reactions, but it can also be applied to several phenomena, 
such as a decay of unstable states, including nuclear fission, 
and chemical reactions. These will be interesting future works. 

\section*{Acknowledgements}
We thank D. Lacroix, D. Regnier, J. Randrup and A. Ohnishi 
for useful discussions.
We are grateful to the Yukawa Institute for Theoretical Physics (YITP), 
Kyoto University. Stimulated discussions during the YITP international 
molecular-type 
workshop ``Nuclear Fission Dynamics 2019'' were 
useful to complete this work. 
This work was supported in part by the Graduate Program on Physics for 
the Universe at Tohoku university, and in part by 
JSPS KAKENHI Grant Number JP19K03861.

\end{document}